\documentclass[aps,prl,twocolumn,nofootinbib]{revtex4}
\usepackage[hypertex]{hyperref}
\usepackage{graphicx} 
\usepackage{amsbsy}   

\def\simgt{\mathrel{\lower2.5pt\vbox{\lineskip=0pt\baselineskip=0pt
           \hbox{$>$}\hbox{$\sim$}}}}
\def\simlt{\mathrel{\lower2.5pt\vbox{\lineskip=0pt\baselineskip=0pt
           \hbox{$<$}\hbox{$\sim$}}}}

\def\vev#1{\left\langle#1\right\rangle}
\def\stacksymbols #1#2#3#4{\def\theguybelow{#2}
    \def\vp{\lower#3pt}
    \def\sp{\baselineskip0pt\lineskip#4pt}
    \mathrel{\mathpalette\intermediary#1}}

\def\intermediary#1#2{\vp\vbox{\sp
     \everycr={}\tabskip0pt
     \halign{$\mathsurround0pt#1\hfil##\hfil$\crcr#2\crcr
              \theguybelow\crcr}}}

\def\gapproxeq{\stacksymbols{>}{\sim}{2.5}{.2}}
\def\lapproxeq{\stacksymbols{<}{\sim}{2.5}{.2}}

\def\beq{\begin{equation}}
\def\eeq#1{\label{#1}\end{equation}}
\def\eeq{\end{equation}}
\def\beqa{\begin{eqnarray}}
\def\eeqa#1{\label{#1}\end{eqnarray}}
\def\eeqan{\end{eqnarray}}

\begin{document}

\pagestyle{plain}

\title{Evolving Dark Energy with {\boldmath $w \neq -1$}}

\author{Lawrence J.~Hall, Yasunori Nomura, Steven J.~Oliver}
\affiliation{Department of Physics, University of California, Berkeley, and\\
  Theoretical Physics Group, Lawrence Berkeley National Laboratory,
  Berkeley, CA 94720, USA}


\begin{abstract}
Theories of evolving quintessence are constructed that generically lead to 
deviations from the $w = -1$ prediction of non-evolving dark energy.  The 
small mass scale that governs evolution, $m_\phi \approx 10^{-33}~{\rm eV}$, 
is radiatively stable, and the ``Why Now?'' problem is solved.  These results 
rest crucially on seesaw cosmology: in broad outline, fundamental physics 
and cosmology can be understood from only two mass scales, the weak 
scale, $v$, and the Planck scale, $M$.  Requiring a scale of dark energy 
$\rho_{\rm DE}^{1/4}$ governed by $v^2/M$, and a radiatively stable 
evolution rate $m_\phi$ given by $v^4/M^3$, leads to a distinctive form 
for the equation of state $w(z)$ that follows from a cosine quintessence 
potential.  An explicit hidden axion model is constructed.  Dark energy 
resides in the potential of the axion field which is generated by a new 
QCD-like force that gets strong at the scale $\Lambda \approx v^2/M \approx 
\rho_{\rm DE}^{1/4}$.  The evolution rate is given by a second seesaw that 
leads to the axion mass, $m_\phi \approx \Lambda^2/f$, with $f \approx M$.
\end{abstract} \pacs{} \maketitle


{\bf 1. Introduction} --- The dominant energy density in the universe has 
negative pressure, causing a recent acceleration in the expansion of the 
universe~\cite{Perlmutter:1998np}, and is known as dark energy.  What is 
the physical picture for this unusual fluid?  How can the size of its energy 
density, $\rho_{\rm DE} \approx (10^{-3}~{\rm eV})^4$, be understood and 
how can the underlying physics be probed?

One interpretation of dark energy is in terms of a parameter $\Lambda$ 
that determines a fixed energy and pressure for the vacuum --- Einstein's 
cosmological constant.  While the size of the small mass scale, 
$10^{-3}~{\rm eV}$, has not been derived from a more basic theory, 
it could, perhaps, be broadly understood from mild anthropic 
arguments~\cite{Weinberg:1987dv}.  Alternatively, dark energy may be 
associated with the dynamics of some scalar field which is uniform in 
space, $\phi(t)$~\cite{Peebles:1987ek,Weiss:1987xa}.  Perhaps the simplest 
possibility is that the potential for this field, $V(\phi)$, is determined 
by the single meV mass scale together with dimensionless couplings 
of order unity.  Such theories of ``acceleressence'' are easy to 
construct~\cite{Chacko:2004ky}, including radiative stability of the 
meV scale, but lead to generic observational consequences for dark 
energy identical to those from a cosmological constant.  Since the time 
scale for $\phi$ evolution, ${\rm meV}^{-1} \approx 10^{-12}~{\rm sec.}$, 
is much less than the present age of the universe, $t_0 \approx 
10^{18}~{\rm sec.}$, the field has already evolved to a local minimum
of the effective potential.

An equation of state differing from that of the cosmological constant 
results if the time scale for $\phi$ evolution is of order $t_0$.  Taylor 
expanding the potential $V(\phi)$ about $\phi_0$, todays value of the field, 
such theories of quintessence require a dynamical scale
\begin{equation}
  m_\phi = \sqrt{V^{''}(\phi_0)} \approx H_0 \approx 10^{-33}~{\rm eV}. 
\label{eq:mphi}
\end{equation}
The appearance of such a low mass scale immediately raises questions.  Can 
such a mass scale be protected from radiative corrections?  If a mechanism 
can be found to stabilize $m_\phi$ to $10^{-33}~{\rm eV}$, then presumably 
it could protect much smaller scales as well, corresponding to a quintessence 
theory where $\phi$ is effectively frozen today, with $V(\phi)$ acting as 
a cosmological constant.  It is for these reasons, perhaps, that there is 
a theoretical expectation that $w = p/\rho$ will be found to be $-1$ and 
time independent.  However, this expectation ignores the constraints that 
will be placed on any theory of dark energy by requiring that it solves the 
radiative stability constraints and the ``Dark Energy Why Now?'' problem.

Why do we live during an era when the energy densities in dark matter and 
dark energy are comparable?  This is the well-known ``Dark Energy Why Now?'' 
problem.  Particle physics provides a simple solution to this problem, at 
least at the order of magnitude level~\cite{Arkani-Hamed:2000tc}.  Particle 
physics can be broadly understood in terms of two fundamental mass scales: 
the reduced Planck scale, $M \approx 10^{18}~{\rm GeV}$, and the electroweak 
scale $v \approx 10^3~{\rm GeV}$.  There is an induced seesaw scale, 
$v^2/M$, that is also of great interest.  Both the Planck and weak eras 
were undoubtedly interesting periods in the evolution of the universe, 
and we expect that the seesaw era, with a temperature of order $v^2/M 
\approx 10^{-3}~{\rm eV} \approx 10~{\rm K}$, will also be an interesting 
epoch.  It is significant that the observed background radiation temperature 
is within an order of magnitude of this value --- we do indeed live during 
the seesaw era.  During this era, at a temperature of $v^2/M$, any particle 
species, or fluid, with an energy density that depends parametrically on 
$M$ and $v$ as $(v^2/M)^4$ would be expected to contribute a significant 
fraction to the energy density of the universe.  The ``Dark Energy Why 
Now?'' problem is solved if theories for dark energy and dark matter can 
be constructed that have this parametric form for their energy densities.

If an evolving quintessence field gives a significant departure of $w$ from 
$-1$, there is a ``Quintessence Why Now?'' problem: why do we live during 
an era when the $\phi$ field is just starting to evolve?  From (\ref{eq:mphi}) 
this becomes: why is $m_\phi \approx H_0 \approx 10^{-33}~{\rm eV}$ and not 
much smaller?  In seesaw cosmology the present value of the Hubble parameter 
is given by $H_0 \approx v^4/M^3$. Once again, seesaw cosmology allows 
a solution to an otherwise intractable problem: the dynamical mass scale 
causing the evolution of $\phi$ must be given parametrically by 
\begin{equation}
  m_\phi \approx v^4/M^3. 
\label{eq:mphiseesaw}
\end{equation}
In quintessence theories, we can expect to observe deviations from $w = -1$ 
if the mass scales in $V(\phi)$ are appropriately related to the 
electroweak scale $v$.  If the mass parameters of $V(\phi)$ are not related 
to those of known particle physics, it does not appear possible to answer 
this problem, except perhaps with anthropic arguments~\cite{Garriga:2003hj}.

In this letter we study quintessence in the seesaw cosmology framework. 
We exhibit a large class of theories that are radiatively stable and 
automatically solve the ``Quintessence Why Now?'' problem.  It is much more 
constraining to also solve the usual ``Dark Energy Why Now?'' problem, and 
we are led to a particular class of axion-like models.


{\bf 2. Radiative Stability and Deviations from {\boldmath $w = -1$}} 
--- From a particle physics perspective, the potential $V(\phi)$ is 
extraordinarily flat~\cite{Kolda:1998wq}.  Supersymmetry is commonly used 
to protect scalar masses at the mass scale $v$, and can even protect certain 
scalars to $v^2/M$ as needed for acceleressence theories, but this is far 
from the desired scale of (\ref{eq:mphiseesaw}).  Factors of $1/16\pi^2$ 
from quantum loops are hardly likely to help.  We are thus led to introduce 
a small parameter $\mu^4$ which explicitly breaks the shift symmetry 
$\phi \rightarrow \phi + c$:
\begin{equation}
  V(\phi) = \mu^4 F\Bigl(\frac{\phi}{f}\Bigr) + {\rm h.c.}
\label{eq:V}
\end{equation}
The dimensionless function $F$ is arbitrary, and for simplicity we have 
assumed that it depends on only a single dimensionful parameter $f$. 
Throughout, we assume that the approximate global symmetries of interest 
are sufficiently protected from any corrections involving non-perturbative 
quantum gravity.  In general $F$ depends on many dimensionless parameters 
that are taken to be of order unity.  We assume that the initial value 
of $\phi$ is of order $f$, and that, since today $\phi$ is at most 
slowly evolving, $\phi_0$ is also of order $f$.  The observed size of 
$\rho_{\rm DE}$ then implies that $\mu$ must be taken of order the meV 
scale.  To solve the ``Dark Energy Why Now?'' problem we will later seek 
theories that lead to $\mu \approx v^2/M$.  In the limit that $\mu^4 
\rightarrow 0$, shift symmetry requires the potential to vanish.  Hence 
all radiative corrections to $V$ are proportional to $\mu^4$ --- the 
potential is radiatively stable.  A pseudo-Goldstone boson provides 
a well-known example of quintessence with radiative stability, in which 
case $F$ is a cosine~\cite{Weiss:1987xa,Frieman:1995pm}.

The dynamical mass scale for $\phi$ evolution is $m_\phi \approx \mu^2/f$. 
Once the dark energy dominates, the Friedmann equation gives $H_0 \approx 
\sqrt{G \rho} \approx \mu^2/M$, leading to 
\begin{equation}
  m_\phi \approx \frac{M}{f} H_0.
\label{eq:mphistab}
\end{equation}
The slow role condition becomes $f \gapproxeq M$.  In the framework of seesaw 
cosmology, there are only two fundamental mass scales $M$ and $v$, and so 
we must choose $f \approx M$.  This gives $m_\phi \approx H_0$ so that the 
``Quintessence Why Now?'' problem is solved; the slow roll condition is lost 
during the present era and deviations from $w=-1$ are generically expected. 
With $f \approx M$, one immediately finds $m_\phi \approx \mu^2/M$, and 
with $\mu \approx v^2/M$ the double seesaw $m_\phi \approx (v^2/M)^2/M$ leads 
to the desired relation (\ref{eq:mphiseesaw}).  To explain why $\mu \approx 
v^2/M$, and to be more precise about the prediction for $w(z)$, we must 
address the ``Dark Energy Why Now?'' problem. 


{\bf 3. A Dynamical \mbox{\boldmath $\mu^4$}} --- As long as $\mu^4$ appears 
as an independent free parameter of the theory, the ``Dark Energy Why Now?'' 
problem will remain unsolved.  To make progress, $\mu^4$ must itself be 
understood to arise dynamically $\mu^4 \rightarrow \lambda G(\chi)$, with 
$G$ a product of fields $\chi$ which may include scalars and fermions. 
A simple example is $G = \chi^4$, with $\chi$ a scalar.  The introduction of 
propagating fields $\chi$ changes the radiative structure of the theory --- 
the parameter which explicitly breaks the shift symmetry on $\phi$ is now 
$\lambda$, which we take to be dimensionless and order unity.  For example, 
integrating over internal $\chi$ fields induces a radiative correction to 
the potential at order $|\lambda|^2$: $\Delta V(\phi) = |\lambda|^2 M^4 
|F(\phi/f)|^2$, giving a $\phi$ mass of order $\lambda M^2/f$.  Indeed, 
treating $\lambda$ as the spurion for shift symmetry breaking, such a term 
cannot be forbidden.  By making $\mu^4$ dynamical, $m_\phi$ is generically 
changed from order $H_0$ to order $\lambda M$!  Even if the loop integrals 
are cutoff by supersymmetry, $m_\phi$ can only be protected to $v^2/M$, 
sufficient for acceleressence, but very far from the requirements of 
dynamical quintessence.  This disastrous radiative correction, however, is 
easily removed by taking $F = e^{i \phi/f}$.  In this case the potential is 
periodic, and $\phi$ is understood to be the pseudo-Goldstone boson of some 
symmetry $U(1)_\phi$ that is spontaneously broken at scale $f$ near the 
Planck scale.  Our potential $V$ then takes the form
\begin{equation}
  V(\phi, \chi) = \lambda G(\chi) \; e^{i \frac{\phi}{f}} + {\rm h.c.}
\label{eq:V2}
\end{equation}

There are other potentially problematic radiative corrections to the 
potential for $\phi$ from diagrams involving $\chi$ loops.  For example, 
if $\chi$ is a scalar and $G = |\chi|^4$, then there are radiative 
corrections at order $\lambda$ in which the four $\chi$ fields are 
contracted into a two loop diagram.  To avoid such contributions $G$ must 
carry some charge under some symmetry $U(1)_\chi$.  For example, with $\chi$ 
a complex scalar and $G = \chi^4$ it is not possible to contract the $\chi$ 
fields into loops as long as there are no other interactions which violate 
$U(1)_\chi$.  In such theories the interaction (\ref{eq:V2}) explicitly 
breaks one combination of $U(1)_\phi$ and $U(1)_\chi$.  The parameter 
$\mu^4$ is generated by having $\chi$ develop an expectation value $f'$, 
so that $\lambda G \rightarrow \lambda \vev{G} e^{i \phi'/f'} \equiv 
\mu^4 e^{i \phi'/f'}$, giving a potential
\begin{equation}
  V(\phi, \phi') = \mu^4 \cos \left( \frac{\phi}{f} + \frac{\phi'}{f'} \right).
\label{eq:V3}
\end{equation}
To obtain a small value for $\mu^4$, we require $f' \ll f \approx M$.  The 
pseudo-Goldstone boson $\phi' + (f'/f)\phi$ then acquires a mass $\mu^2/f' 
\gg H_0$, while $\phi - (f'/f)\phi'$ remains an exactly massless Goldstone 
boson.  Therefore, at this point there is no candidate for the dynamical 
quintessence field.

The situation is radically altered if some additional explicit symmetry 
breaking interaction is added, $\widetilde{V}$, giving a mass to $\phi'$ 
that is $\gapproxeq \mu^2/f'$.  In this case the determinant of the 
pseudo-Goldstone-boson mass matrix no longer vanishes, so that the previously 
massless Goldstone boson acquires a mass from (\ref{eq:V3}): $m_\phi = 
\mu^2/f \approx H_0$.  Thus dynamical quintessence theories naturally emerge 
from theories having the explicit symmetry breaking structure
\begin{equation}
  U(1)_\phi \times U(1)_\chi 
    \stackrel{\widetilde{V}}{\rightarrow}
  U(1)_{\phi + \chi} 
    \stackrel{\lambda}{\rightarrow} 
  0,
\label{eq:ESB}
\end{equation}
with the mass of the dark energy field emerging at the final stage of 
explicit symmetry breaking. 

The form of $\widetilde{V}$ is itself highly constrained, since radiative 
corrections involving both $\lambda$ and $\widetilde{V}$ must not introduce 
further operators that give a large mass to $\phi$.  To avoid this, 
the explicit symmetry breaking parameter in $\widetilde{V}$ should be 
dimensionful.  For example, the case of $G = \chi^4$ and $\widetilde{V} 
= \eta \chi^4 + {\rm h.c.}$ clearly does not work.%
\footnote{An important question is whether theories of the form 
$m_{\nu_{ij}} \nu_i \nu_j e^{i \phi_{ij}/f_{ij}}$ lead to acceptable 
potentials for dark energy once the three neutrino fields $\nu_i$ are 
integrated out.  If $m_{\nu_{ij}}$ are treated as parameters, one obtains 
a potential of the form of (\ref{eq:V3}) with $\mu$ identified as 
$m_\nu$~\cite{Frieman:1995pm}.  This would be a very interesting 
understanding of the size of dark energy.  However, the simplest such 
theories do not work: the neutrino mass is not a parameter but depends 
on electroweak symmetry breaking $m_\nu = m_\nu(h)$, and radiative 
corrections above the weak scale with internal Higgs fields, $h$, 
destroy the radiative stability of the potential.  The schizon models 
of~\cite{Hill:1988bu} avoid this by introducing multiple Higgs doublets 
at the weak scale. But, even in this case, the mass parameters that 
mix the various Higgs doublets must be set to the weak scale by hand 
--- they cannot arise from vacuum expectation values of other fields. 
The successful supersymmetric prediction for the weak mixing angle 
is also destroyed.}


{\bf 4. Hidden Axions and Seesaw Cosmology} --- To illustrate these 
ideas, and to see how seesaw cosmology can solve the ``Dark Energy 
Why Now?'' problem, we consider models with an axion in a hidden 
sector. Quintessence axions have been considered previously for dark 
energy~\cite{Nomura:2000yk,Kim:1998kx}, but not in the context of 
seesaw cosmology.

The general idea is as follows.  Suppose that the fundamental scale of 
supersymmetry breaking in nature is of order of the TeV scale, $v$.  Any 
sector of the theory that feels this supersymmetry breaking only indirectly 
via gravity mediation will have an effective scale of supersymmetry breaking 
at the seesaw scale $\widetilde{m} = v^2/M$. We suppose that such a hidden 
sector has a supersymmetric QCD-like gauge interaction acting on chiral 
superfields $Q$ and $Q^c$.  Supersymmetry breaking leads to the corresponding 
squarks and gluinos acquiring a mass of order $\widetilde{m}$, changing the 
beta function for the gauge coupling and triggering strong dynamics at a scale 
$\Lambda$ not far below $\widetilde{m}$.  A simple example for this behavior 
arises if the hidden sector is in a conformal window above $\tilde{m}$. 
We assume that supersymmetry breaking also triggers a mass term for the 
quarks.  If this sector has a Peccei-Quinn symmetry spontaneously broken 
at $f$ near the Planck scale, then the interaction between the axion, 
$\phi$, and the quarks at the scale $\Lambda$ has the form
\begin{equation}
  {\cal L}_{\rm ax} = m_q\, q q^c\; e ^{i \frac{\phi}{f}} + {\rm h.c.}
\label{eq:coup}
\end{equation}
so that, comparing with (\ref{eq:V2}), $\lambda G = m_q q q^c$.  The 
$U(1)_\phi$ symmetry is the Peccei-Quinn symmetry, $U(1)_{\rm PQ}$, and 
is broken near the Planck scale, while the $U(1)_\chi$ symmetry is the 
axial $U(1)$ symmetry, $U(1)_A$, carried by the quark bilinear $q q^c$. 
The interaction (\ref{eq:coup}) explicitly breaks $U(1)_{\rm PQ} \times
U(1)_A$ to the diagonal subgroup.  We assume that the mass of at least one 
quark flavor in (\ref{eq:coup}) is $\lapproxeq \Lambda$, so that a condensate 
forms, $\vev{q q^c} \approx \Lambda^3 e^{i \eta'/\Lambda}$, generating the 
potential (\ref{eq:V3}) with $\phi'$ becoming the hidden sector $\eta'$ 
and $f' = \Lambda$. 

The additional explicit symmetry breaking necessary for a naturally light 
quintessence field, $\widetilde{V}$ in (\ref{eq:ESB}), is automatic: it is 
the gauge anomaly that breaks $U(1)_A$ giving the $\eta'$ a mass of order 
$\Lambda$.  Since this explicit symmetry breaking comes from an anomaly 
and involves the scale $\Lambda$, unlike dimensionless symmetry breaking 
parameters, it does not lead to further radiative instabilities of the mass 
of the dark energy field.  The axion field $\phi$ is the dark energy field, 
and obtains a mass from the potential (\ref{eq:V3}) with $\mu^4 \approx 
m_q \Lambda^3$.  Since $\Lambda$ and $m_q$ are both close to $\widetilde{m}$, 
the scale $\mu$ is given by the seesaw $\mu \approx \widetilde{m} \approx 
v^2/M$, solving the ``Dark Energy Why Now?'' problem.  The double seesaw
\begin{equation}
  m_\phi \approx \frac{\mu^2}{f}, \quad\quad \mu \approx \frac{v^2}{M},
\label{eq:doubless}
\end{equation}
then leads to the desired result (\ref{eq:mphiseesaw}) for a seesaw 
cosmology solution of the ``Quintessence Why Now?'' problem.

It is straightforward to write a complete set of interactions for the above 
hidden sector.  As an example, consider the supersymmetric interaction 
Lagrangian
\begin{eqnarray}
  && {\cal L}_{\rm int} = \int\! d^2 \theta 
    \left( X (S \bar{S} - f^2) + {Z \over M} W^\alpha W_\alpha \right)
\nonumber \\
  && + \int\! d^4 \theta 
    \left( {Z^\dagger \over M} {S \over M} Q Q^c 
      + {Z^\dagger Z \over M^2}(Q^\dagger Q + Q^{c \dagger} Q^c) \right),
\label{eq:Lint}
\end{eqnarray}
where all coupling constants, color and flavor indices have been omitted. 
The chiral superfield $Z$ is the spurion for supersymmetry breaking with 
$F_Z/M = \widetilde{m} \approx v^2/M$.  The interactions of (\ref{eq:Lint}) 
possess $U(1)_{\rm PQ} \times U(1)_B \times U(1)_R$ symmetry, where $U(1)_B$ 
is the baryon symmetry acting on $Q$ and $Q^c$ and $U(1)_R$ the $R$ symmetry 
under which $\tilde{q}$ and $\tilde{q}^c$ are neutral.  We assume that 
$U(1)_R$ is explicitly broken elsewhere in the theory, and $U(1)_B$ plays 
no role in our analysis.  The two relevant symmetries are then $U(1)_{\rm PQ}$ 
and the axial $U(1)_A$ symmetry on $Q$ and $Q^c$.  These are $U(1)_\phi$ 
and $U(1)_\chi$, respectively, and the two explicit symmetry breakings of 
(\ref{eq:ESB}) are provided by the gauge anomaly and by the interaction 
$Z^\dagger S Q Q^c$, respectively.  The chiral field $X$ drives the 
spontaneous breaking of $U(1)_{\rm PQ}$ symmetry, giving $\vev{S} = 
f e^{i \phi/f}$.  All hidden sector superpartners obtain a mass of order
$\widetilde{m}$ through interactions with $Z$.  On inserting $F_Z$ and 
$\vev{S}$ into the interaction $Z^\dagger S Q Q^c$, a supersymmetric mass 
term for the quarks is generated, which includes the desired interaction 
of (\ref{eq:coup}).  We do not include a phase for the pseudo-Goldstone boson 
of $U(1)_R$ because it acquires a sufficiently large mass from elsewhere. 
We assume that the flat direction associated with the real parts of
$S$ and $\bar{S}$ can be sufficiently lifted.

There are many alternative models. For example, the hidden 
sector could be a copy of the supersymmetric standard model coupled to 
a Planck scale axion.


{\bf 5. Equation of State Predictions} ---  
\begin{figure}
\begin{center}
  \includegraphics[width=8.6cm]{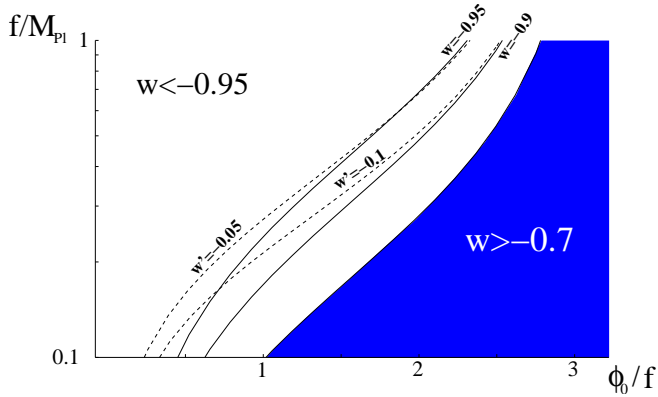}
\caption{Contours for $w$ and $w'$.}
\label{fig:ww'}
\end{center}
\end{figure}
\begin{figure}
\begin{center}
  \includegraphics[width=8.6cm]{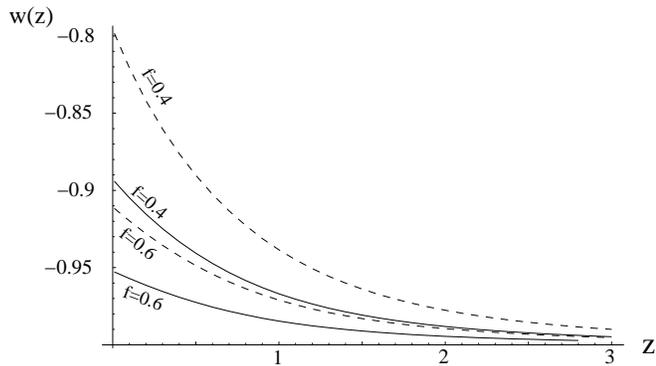}
\caption{$w(z)$ for 4 choices of $(f/M_{\rm Pl}, \phi_0/f)$.  The solid 
lines are $\phi_0/f = 0.6\pi$ while the dashed lines have $\phi_0/f 
= 0.7\pi$.  The lines are labeled by their value of $f/M_{\rm Pl}$.}
\label{fig:wz}
\end{center}
\end{figure}
The class of quintessence theories we have introduced, having a radiatively 
stable potential resulting from a shift symmetry and a solution to the 
``Dark Energy Why Now?'' problem via $\mu \approx v^2/M$, leads to a 
potential of the form $V = \mu^4 \cos(\phi/f)$, with $f \approx M$. 
Thus the dark energy and its cosmological evolution is described by three 
parameters: $\mu^4$, $f$ and $\phi_0$.  We choose to determine $\mu^4$ 
from the observed size of $\rho_{\rm DE}$, and display predictions in 
the ($f/M_{\rm Pl}$, $\phi_0/f$) plane, where $M_{\rm Pl} = 1.2 \times 
10^{19}~{\rm GeV}$.  In Figure~\ref{fig:ww'} contours are drawn for 
$w = -0.7, -0.9$ and $-0.95$ and also for $w' = -0.1$ and $-0.05$, 
where $w' = dw/dz|_{z=0}$.  There is a sizable region of allowed 
parameter space having deviations from $w = -1$ observable at future 
experiments~\cite{Aldering:2002dp}.  In Figure~\ref{fig:wz} the redshift 
dependence of the equation of state parameter, $w(z)$, is shown for four 
representative values of $(f/M_{\rm Pl}, \phi_0/f)$.  Recent evolution 
can be quite rapid and is determined by the cosine form of the potential.


\acknowledgments{This work was supported in part by the DOE under contracts 
DE-FG02-90ER40542 and DE-AC03-76SF00098 and in part by NSF grant PHY-0098840. 
The work of Y.N. was also supported by NSF grant PHY-0403380 and by a DOE 
OJI award.}

\end{document}